# Solving Cooperative Reliability Games

Yoram Bachrach†, Reshef Meir‡, Michal Feldman§, Moshe Tennenholtz‡

†Microsoft Research, Cambridge, UK
‡Microsoft Research, Hertzelia, Israel
§School of Business Administration, Hebrew University and Microsoft research, Herzliya, Israel

## Abstract

Cooperative games model the allocation of profit from joint actions, following considerations such as stability and fairness. We propose the reliability extension of such games, where agents may fail to participate in the game. In the reliability extension, each agent only "survives" with a certain probability, and a coalition's value is the probability that its surviving members would be a winning coalition in the base game. We study prominent solution concepts in such games, showing how to approximate the Shapley value and how to compute the core in games with few agent types. We also show that applying the reliability extension may stabilize the game, making the core non-empty even when the base game has an empty core.

## 1 Introduction

Consider a communication network connecting a source and target vertices, where each link is controlled by a self-interested agent. Any link may fail, and if such failures result in the elimination of all paths between the source and target, information cannot be sent between them. Consider a planner aiming to maximize the probability of allowing communication between the source and target. Each link has its own probability of failure, and together with the network structure they determine the probability of connectivity between the source and target. When the planner is only allowed to use a certain subset of the links, we may consider links outside this set as failed links. Given a reward for achieving such connectivity, what is a reasonable way of allocating this reward between the agents? Which links are the most critical?

This domain can be modeled as a cooperative game, which we call the network reliability game. In this game we are uncertain which links would survive, but link failures are independent. The value of a link subset, called a coalition, is the probability of achieving connectivity using only these links. Game theoretic solutions find ways of allocating the reward between the agents, under considerations of fairness and stability. The most prominent solution aiming at fairness is the *Shapley value* [19], which is a *measure of the criticality* of each edge to achieving connectivity. The *core* [13] is the most common solution considering stability, and contains all allocations where no subset of agents is incentivized to defect and form an alternative network. Applying these solutions requires finding tractable algorithms for computing them . Unfortunately, even in domains where determining the value of a subset of agents is easy, computing these solutions may be hard. Further, the problem of computing the value of a coalition in the network reliability game is the famous *network reliability problem*, which is known to be computationally hard [16], making it even harder to apply the required solutions.

**Our contribution** We propose the *reliability extension* of cooperative games, providing a general framework for studying the effects of uncertain failures. We focus on techniques for solving such games. We show how to approximate the Shapley value in such games using sampling, and how to accurately compute the core in games with few agent types. We also show that in simple games (such as our network game), reducing the reliability of agents can only expand the core, surprisingly making the game *more* stable. Section 2 provides some definitions, Section 3 considers the Shapley value in network settings, Section 4 extends the results for general domains and Section 5 examines the core. We conclude in Section 6.

## 2 Preliminaries

A transferable utility cooperative game is composed of a set of $n$ agents, $N$, and a characteristic function

$v : 2^N \to \mathbb{R}$ mapping any subset (coalition) of the agents to a real value indicating the total utility these agents achieve together. By convention $v(\emptyset) = 0$. A game is *monotone* if for all coalitions $C' \subset C$ we have $v(C') \leq v(C)$. A game $H = \langle N, v \rangle$ is *simple*, if $v$ only gets values of 0 or 1 ($v : 2^N \to \{0, 1\}$). We say coalition $C \subset N$ *wins* if $v(C) = 1$, and say it *loses* if $v(C) = 0$. An agent $i$ is *critical* in a winning coalition $C$ if her removal from that coalition would make it a losing coalition: $v(C) = 1$, $v(C \setminus \{i\}) = 0$. A game is *convex* if for any $A, B \subseteq N$ we have $v(A \cup B) \geq v(A) + v(B) - v(A \cap B)$.

The characteristic function defines what gains a coalition achieves. Cooperative game theory provides solution concepts that define how the participants might agree to *distribute* the gains. An *imputation* $\mathbf{p} = (p_1, \ldots, p_n)$ is a division of the gains among the agents, where $p_i \geq 0$ and $\sum_{i=1}^n p_i = v(N)$. The value $p_i$ is the payoff of agent $i$, and the payoff of a coalition $C$ is $p(C) = \sum_{i \in C} p_i$.

**The core**  A basic requirement for a good imputation is that for any agent $i \in C$, we have $p_i \geq v(\{i\})$, otherwise $i$ is better off working alone rather than as a part of $C$ (which only offers her $p_i$). Similarly, a coalition $B$ *blocks* imputation $\mathbf{p}$ if $p(B) < v(B)$, since $B$'s members can defect, derive the gains $v(B)$, give each member $i \in B$ its previous share $p_i$, and still some utility remains, so each member can get more. If a blocked imputation is chosen, the coalition is unstable. The *core* is the set of all imputations that are not blocked by any coalition. That is, if $\mathbf{p} \in core(H)$, then for any coalition $C$ we have $p(C) \geq v(C)$. The core is the prominent solution concept focusing on stability [13], but it may be empty or contain more than one imputation.

**The Shapley value**  The Shapley value focuses on *fairness* rather than on stability, by evaluating the contribution of each agent to the grand coalition $N$. Further, it is the unique imputation that fulfills important fairness axioms [19]. The Shapley value relies on the notion of a marginal contribution of an agent in a permutation, the amount of additional utility generated when that agent joins the coalition of her predecessors in the permutation. We denote by $\pi \in \mathcal{S}_n$ a permutation of the agents, so $\pi : \{1, \ldots, n\} \to \{1, \ldots, n\}$ and $\pi$ is onto. Denote by $\Gamma_i^\pi$ the predecessors of $i$ in $\pi$, so $\Gamma_i^\pi = \{j | \pi(j) < \pi(i)\}$. Agent $i$'s marginal contribution in the permutation $\pi$ is $m_i^\pi = v(\Gamma_i^\pi \cup \{i\}) - v(\Gamma_i^\pi)$. Note that in a simple game an agent has a marginal contribution of 1 in the permutation $\pi$ iff it is critical for the coalition $\Gamma_i^\pi$. The Shapley value of an agent is her marginal contribution averaged across all possible agent permutations.

**Definition 1.** *The Shapley value is the imputation* $(\phi_1(v), \ldots, \phi_n(v))$ *where*

$$\phi_i(v) = \frac{1}{n!} \sum_{\pi \in \mathcal{S}_n} m_i^\pi = \frac{1}{n!} \sum_{\pi \in \mathcal{S}_n} \left( v\left(\Gamma_i^\pi \cup \{i\}\right) - v\left(\Gamma_i^\pi\right) \right)$$

One interpretation of "power" is an agent's *a priori probability* of being critical in determining the outcome of the game. Thus the Shapley value measures an agent's power to affect the outcome of the game.

## 3  Network Reliability Games

Consider the network reliability domain discussed in Section 1. Suppose the the planner obtains a reward of \$1 for achieving $s - t$-connectivity. We can define the value of a set $C$ of edges as the planner's expected reward when she is only allowed to use the edges in $C$. Since the reward is \$1, the value of $C$ is the probability of the planner to achieve connectivity when only allowed to use the edges of $C$.

Consider a directed graph $G = \langle V, E \rangle$ with a source $s \in V$ and target $t \in V$, where each edge $e_i$ is controlled by an agent $a_i$ and has a failure probability of $1 - r_i \in [0, 1]$. Thus an edge $e_i$ survives with probability $r_i$. Given a subset $S \subseteq E$ of the edges, we use $\delta_S$ as the characteristic denoting whether $S$ is successful, i.e. whether it allows connectivity between $s$ and $t$:

$$\delta_S = \begin{cases} 1 & \text{if } G_S = \langle V, S \rangle \text{ has a path from } s \text{ to } t \\ 0 & \text{otherwise} \end{cases}$$

The graph's structure determines $\delta_S$, which can be viewed as a function mapping any coalition $S$ to $\{0, 1\}$, denoting whether $S$ is successful or not. We can define a game with a characteristic function $v_{base}$ where agents are the edges, and the value of coalition $S$ is $v_{base}(S) = 1$ if $S$ has a path from $s$ to $t$ and $v_{base}(S) = 0$ otherwise, so $v_{base}(S) = \delta_S$. However, such a definition does not take into account different edge failure probabilities. This will be handled by the relaiablity extension defined next, which we generalize to other set-systems in Section 4.

Suppose the planner can only use the edges $C \subseteq E$. It cannot use edges in $E \setminus C$, and even the edges in $C$ may not survive. Each edge $i \in C$ survives with probability $r_i$, so from the original edges $E$ we get a set $C' \subseteq C$ of surviving edges. Denote the probability that the edges surviving from $C$ are *exactly* those in $C'$ as $Pr(C'|C) = \prod_{i \in C'} r_i \cdot \prod_{j \in C \setminus C'} (1 - r_j)$. The success of the surviving edges is a random variable, and the probability of the surviving edges to be successful is:

$$\sum_{C' \subseteq C} Pr(C'|C) \cdot \delta_{C'} = \sum_{C' \subseteq C} \left( \prod_{i \in C'} r_i \prod_{j \in C \setminus C'} (1 - r_j) \right) \cdot \delta_{C'}.$$

Thus, the above expression is the probability that the planner would achieve its goal of sending a message from $s$ to $t$ when it is only allowed to use the edges $C \subseteq E$.[1] Any subset $C \subseteq E$ the planner is allowed to use results in a different probability of connectivity between $s$ and $t$. We define a game where the edges are the agents, and the characteristic function $v : 2^E \to \mathbb{R}$ maps any subset of edges $C$ to the probability that the planner would achieve connectivity between $s$ and $t$ when only the edges $C$ are at its disposal.

**Definition 2 (The Network Reliability Game).** *Given a graph $G = \langle V, E \rangle$ and survival probabilities $r_i$ for each $e_i \in E$, the network reliability game is a game over the agents $E$ with the characteristic function:*

$$v(C) = \sum_{C' \subseteq C} \left( \prod_{i \in C'} r_i \cdot \prod_{j \in C \setminus C'} (1 - r_j) \right) \cdot \delta_{C'}.$$

As an example consider a network with one edge $e$ connecting the source to the target directly, and a parallel path $C$ of 3 edges connected serially and forming a path between the source and target. Assume each link has a survival probability of $\frac{1}{2}$. Consider the coalition $C$ of the serial edges only. To achieve connectivity all three must survive, an event whose probability is $\frac{1}{2^3} = \frac{1}{8}$. The value of this coalition $C$ is thus $v(C) = \frac{1}{8}$. Adding the direct link $e$ to $C$ increases this probability to $v(C \cup \{e\}) = 1 - (\frac{7}{8} \cdot \frac{1}{2})$.

The edges in the network may be unequal in their criticality in achieving connectivity between $s$ and $t$. Some edges have little influence on the probability of achieving the goal because of their location in the network. Others have little influence because their survival probability is low, so including them is unlikely to change the outcome. How should we quantify all these effects to find reliability problems in the network?

A key component in such an analysis is the marginal contribution of an edge in various contexts. One definition of a "context" is a permutation of the agents (edges). Using this definition we can use the Shapley value to rank edges by their impact on the probability of attaining the goal. Previous work already suggests game theoretic solutions (including the Shapley value) for analyzing network reliability problems [12, 5, 1, 4].

---

[1] When the planner obtains one unit of reward for achieving connectivity between $s$ and $t$, this expression is also the expected reward it obtains.

However, these approaches do not model different link failure probabilities. Unfortunately, computing solutions such as the Shapley value are computationally hard in many domains. One way to overcome this difficulty is using approximation algorithms, such as sampling various contexts and computing an agent's contribution in each. The average contribution an agent has in the sampled contexts is used to estimate its average marginal contribution across all contexts. Computing $m_i^\pi = v(\Gamma_i^\pi \cup \{i\}) - v(\Gamma_i^\pi)$, the contribution in context $\pi$, requires computing the values of two coalitions. In the case of the network reliability game, this is the famous *network reliability* problem, known to be #P-hard [16]. Thus, a naïve sampling approach is intractable for this problem.

Although solving *network reliability* exactly is hard, this problem is approximable through sampling. We show how to compute the Shapley value for the network reliability game using a sampling approach. The value of a coalition $C \subseteq N$ is given by:

$$v(C) = \sum_{S \subseteq C} \delta_S Pr(S|C) = \sum_{S \subseteq C} \delta_S \prod_{i \in S} r_i \prod_{j \in C \setminus S} (1 - r_j)$$

The contribution of $x \in N$ in permutation $\pi$ is:

$$m_i^\pi = v(\Gamma_i^\pi \cup \{i\}) - v(\Gamma_i^\pi)$$
$$= \sum_{S \subseteq \Gamma_i^\pi \cup \{i\}} \delta_S Pr(S|\Gamma_i^\pi \cup \{i\}) - \sum_{S \subseteq \Gamma_i^\pi} \delta_S Pr(S|\Gamma_i^\pi)$$

We examine $q_i = \sum_{S \subseteq \Gamma_i^\pi \cup \{i\}} \delta_S Pr(S|\Gamma_i^\pi \cup \{i\})$:

$$q_i = \sum_{S \subseteq \Gamma_i^\pi \cup \{i\}} \delta_S Pr(S|\Gamma_i^\pi \cup \{i\})$$
$$= \sum_{S \subseteq \Gamma_i^\pi \cup \{i\} | i \in S} \delta_S Pr(S|\Gamma_i^\pi \cup \{i\}) + \sum_{S \subseteq \Gamma_i^\pi \cup \{i\} | i \notin S} \delta_S Pr(S|\Gamma_i^\pi \cup \{i\})$$
$$= \sum_{S \subseteq \Gamma_i^\pi} \delta_{S \cup \{i\}} \cdot r_i \cdot Pr(S|\Gamma_i^\pi)$$
$$+ \sum_{S \subseteq \Gamma_i^\pi} \delta_S \cdot (1 - r_r) \cdot Pr(S|\Gamma_i^\pi)$$

We now express $m_i^\pi$ in terms of $v(\Gamma_i^\pi)$:

$$m_i^\pi = r_i \sum_{S \subseteq \Gamma_i^\pi} \delta_{S \cup \{i\}} Pr(S|\Gamma_i^\pi) + (1 - r_i) v(\Gamma_i^\pi) - v(\Gamma_i^\pi)$$
$$= r_i \sum_{S \subseteq \Gamma_i^\pi} \delta_{S \cup \{i\}} Pr(S|\Gamma_i^\pi) - r_i \sum_{S \subseteq \Gamma_i^\pi} \delta_S Pr(S|\Gamma_i^\pi)$$
$$= r_i \sum_{S \subseteq \Gamma_i^\pi} (\delta_{S \cup \{i\}} - \delta_S) Pr(S|\Gamma_i^\pi)$$

The Shapley value of agent $i$ in the game $H$ with characteristic function $v$ can be written as

$$\phi_i(v) = \frac{1}{|N|!} \sum_{\pi \in \mathcal{S}_n} m_i^\pi \quad \text{(Def. 1)}$$

$$= \frac{r_i}{|N|!} \sum_{\pi \in \mathcal{S}_n} \sum_{S \subseteq \Gamma_i^\pi} (\delta_{S \cup \{i\}} - \delta_S) Pr(S|\Gamma_i^\pi)$$

$$= \frac{r_i}{|N|!} \sum_{\pi \in \mathcal{S}_n} \left[ \sum_{S \subseteq \Gamma_i^\pi} (\delta_{S \cup \{i\}} - \delta_S) \prod_{j \in S} r_j \prod_{j' \in \Gamma_i^\pi \setminus S} (1 - r_{j'}) \right]$$

Consider selecting a permutation of $N$ uniformly at random and constructing a random variable by sampling the survival of each of the agents in $\Gamma_i^\pi$:

**Algorithm 1.** $ShapleyApproxBase(v)$:

1. Select $\pi \in \mathcal{S}_n$ uniformly at random from all permutations over $N$.
2. Set $\Gamma_i^\pi$ to be the predecessors of $i$ in $\pi$.
3. Set $l_i = 1$ with probability $r_i$ and $l_i = 0$ with probability $1 - r_i$.
4. If $l_i = 0$ return $Z_i = 0$.
5. For each $j \in \Gamma_i^\pi$ set $l_j = 1$ with probability $r_j$ and set $l_j = 0$ with probability $1 - r_j$.
6. Denote $S = \{j \in \Gamma_i^\pi | l_j = 1\}$ (survivor set)
7. If $\delta_{S \cup \{i\}} - \delta_S = 1$ then set $Z_i = 1$ otherwise set $Z_i = 0$.
8. Return $Z_i$

**Lemma 1.** *Let $Z_i$ be the random variable generated by the above algorithm. Then the expected value of $Z_i$ is the Shapley value of $x$ so: $E(Z_i) = \phi_i(v)$.*

*Proof.* We choose $\pi$ uniformly at random from the set $S_n$ of all edge permutations, so by linearity of expectation:

$E(Z_i) =$

$$\sum_{\pi \in \mathcal{S}_n} \left[ \frac{r_i}{|N|!} \sum_{S \subseteq \Gamma_i^\pi} \prod_{j \in S} r_j \prod_{j' \in \Gamma_i^\pi \setminus S} (1 - r_{j'}) \cdot (\delta_{S \cup \{i\}} - \delta_S) \right]$$

$= \phi_i(v)$,

as required. □

We can repeat the above algorithm $k$ times, to obtain multiple random samples chosen i.i.d from the same distribution. We can then take an average over all samples as a better approximation for the Shapley value $\phi_i(v)$.

**Algorithm 2.** $ShapleyApprox(v, k)$:

1. $Z' = 0$
2. Repeat $k$ times:
   (a) $Z_i = ShapleyApproxBase(v)$
   (b) $Z' = Z' + Z_i$
3. return $Z = \frac{1}{k} \cdot Z'$

The quality of the estimate $Z$ returned by Algorithm 2 depends on $k$, the number of samples used. We use Hoeffding's inequality to evaluate the approximation quality of $Z$ for $\phi_i(v)$. Our goal is to obtain the required number of times $k$ we must sample such that with a high probability $1 - \delta$ we obtain an approximation $Z$ to $\phi_i(v)$ that is within an $\epsilon$ distance from the true value, so $|Z - \phi_i(v)| \leq \epsilon$.

**Theorem 1 (Hoeffding's inequality [14]).** *Let $X_1, \ldots, X_k$ be independent random variables, where all vairables are bounded s.t. $X_t \in [a, b]$, and let $X = \sum_{t=1}^k X_t$. Then the following inequality holds.*

$$\Pr(|X - E[X]| \geq k\epsilon) \leq 2 \exp\left(-\frac{2 k^2 \epsilon^2}{(b-a)^{2k}}\right)$$

We note that all $Z_t$ are numbers between 0 and 1 and that $E[kZ] = E[Z'] = k \cdot \phi_i(v)$. Thus:

$$\Pr(|Z - \phi_i(v)| \geq \epsilon) = \Pr(|kZ - k\phi_i(v)| \geq k\epsilon) \leq 2e^{-2 k \epsilon^2}$$

We next compute the required samples $k$ to bound this probability below $\delta$.

**Theorem 2 (Power Confidence Interval).** *Given a required accuracy $\epsilon > 0$ and confidence $1 - \delta$ the interval $[Z - \epsilon, Z + \epsilon]$, where $Z$ is the result of Algorithm 2, contains $\phi_i(v)$ with probability $1 - \delta$ provided that the number of samples is at least: $k \geq \frac{\ln \frac{2}{\delta}}{2 \epsilon^2}$. Similarly, given a number of samples $k$ and a required confidence of $1 - \delta$, a confidence interval that contains the correct value $\phi_i(v)$ with probability $1 - \delta$ is $\left[Z - \sqrt{\frac{1}{2k} \ln \frac{2}{\delta}}, Z + \sqrt{\frac{1}{2k} \ln \frac{2}{\delta}}\right]$*

*Proof.* We apply Hoeffding's inequality to make bound the error probability at $\delta$ and obtain:

$$Pr(|Z - \phi_i(v)| \geq \epsilon) \leq 2 e^{-2 k \epsilon^2} \leq \delta$$

Extracting the required $\epsilon$ and $k$ we get: $-2 k \epsilon^2 \leq \ln \frac{\delta}{2}$, or alternatively: $\epsilon^2 \geq -\frac{\ln \frac{\delta}{2}}{2 k}$. Thus the relation between the number of samples $k$, the required confidence $\delta$ and the accuracy $\epsilon$ is:

$$\epsilon \geq \sqrt{\frac{1}{2k} \ln \frac{2}{\delta}} \qquad k \geq \frac{\ln \frac{2}{\delta}}{2 \epsilon^2},$$

as required. □

## 4 General Reliability Games

Section 3 has focused on a network domain with a specific goal connectivity goal when edges are susceptible to failures. However, a similar approach can be taken for other network goals, such as achieving a minimal bandwidth or minimizing latency. Similar situations also arise not only in network environments. Generally, consider any domain where a set of agents can cooperate on a task, where some subsets of the agents are successful and can achieve the task, while others cannot do so. Such a domain can be described as a simple game $H$ over the set $N$ of agents, with a characteristic function $v : 2^N \to \{0, 1\}$. We call this game $v$ the *base game*. Now consider the case where each agent $i \in N$ only survives with probability $r_i$ and is eliminated with probability $1 - r_i$. In other words, given that the coalition $C$ is formed, each agent $i \in C$ only survives with probability $r_i$, so the coalition is transformed into the coalition $C' \subseteq C$ of surviving agents. The value achieved by the surviving coalition $C'$ is a random variable. The reliability extension of the base game is a new game, where a coalition $C$'s value is the *probability* that after eliminating agents according to the individual survival probabilities, the surviving agents $C'$ would win the game.[2]

**Definition 3 (The Reliability Extension).** *Given a simple cooperative game $H$ with characteristic function $v : 2^N \to \{0, 1\}$ over the $n$ agents $N$ and the survival probability vector $\mathbf{r} = (r_1, \ldots, r_n)$ (where $0 \leq r_i \leq 1$), we define the reliability game $H^{\mathbf{r}}$ over the same agent set $N$ with the characteristic function $v^{\mathbf{r}}$ as follows.*

$$v^r(C) = \sum_{C' \subseteq C} \alpha^{\mathbf{r}}(C'|C) \cdot v(C'), \qquad (1)$$

*where $\alpha^{\mathbf{r}}(C'|C) = \prod_{i \in C'} r_i \cdot \prod_{j \in C \setminus C'} (1 - r_j)$.*

Note that $\alpha^{\mathbf{r}}(C'|C)$ is $Pr(C'|C)$ according to the survival probability vector $\mathbf{r}$.

The network reliability game is the reliability extension of the base game $v$ where agents are the edges a coalition of edges wins if it contains an $s - t$-path. We denoted the value of coalition $C \subseteq E$ in the base game as $\delta_C$, so $\delta$ is the characteristic function of the base game. We show that our Shapley approximation algorithm can be used for any reliability extension game.

**Corollary 1.** *Let $H$ be a simple cooperative game, with characteristic function $v : 2^N \to \{0, 1\}$, and let $H^{\mathbf{r}} = \langle N, v^{\mathbf{r}} \rangle$ be its reliability extension using the probability vector $\mathbf{r}$. Given an algorithm for computing the base characteristic function $v(C)$ (for any $C \subseteq N$), it is possible to approximate the Shapley $\phi_i(v^{\mathbf{r}})$ of agent $i \in N$ in the reliability extension in polynomial time (in the confidence $\delta$ and accuracy $\epsilon$).*

*Proof.* The analysis in Section 3 used $\delta_S$ to denote the value of a coalition $S$ in the base game, and only assumed that given $S$ one can compute $\delta_S$ in polynomial time. In the network reliability game this required testing for an $s - t$-path in the induced graph. In a general reliability game $v$, $\delta_C$ is the value of the coalition $C$ in the base game, so $\delta_C = v(C)$. Thus, given a polynomial algorithm to compute $v(C)$ we can evaluate $\delta_C$ for any coalition $C$ in polynomial time. Due to Theorem 2, to achieve a confidence $\delta$ and accuracy $\epsilon$ we must run algorithm 2 with a number of samples $k$ that is polynomial in $\epsilon$ and $\delta$. Each sample results in a call to Algorithm 1, which has a polynomial running time when evaluating $\delta_C$ can be done in polynomial time. Thus, the total running time of Algorithm 1 with the required $k$ is also polynomial. □

## 5 The Core of Reliability Games

A natural question is how stability (in terms of strategic behavior) is affected as reliability decreases. While generally various effect are possible, we give a conclusive answer regarding convex and simple games. We show that if a simple or convex base game is stable (i.e. has a non-empty core), then any reliability extension of it is still stable. Further, the core of an extended simple game can be easily characterized.

**Theorem 3.** *Let $H = \langle N, v \rangle$ be a convex game. For any probability vector $\mathbf{r}$ the reliability extension $H^{\mathbf{r}}$ has a non-empty core. Conversely, if $H$ is not convex, then there is at least one extension with an empty core.*

*Proof sketch.* A game $v$ is know to be convex if and only if every subgame of $v$ has a non-empty core [18]. For every $C \in 2^N \setminus \emptyset$, let $p_C$ be a payoff vector in the core of $v|_C$ (the subgame of $v$ that contains only players from $C$). Recall $\alpha^{\mathbf{r}}$ from Definition 3. It is easy to verify that the payoff vector $p^{\mathbf{r}} = \sum_{C \in 2^N \setminus \emptyset} \alpha^{\mathbf{r}}(C|N) p_C$ is in the core of $v^{\mathbf{r}}$.[3] First note that $p^{\mathbf{r}}$ is a valid payoff vector, i.e. that $p^{\mathbf{r}}(N) = v^{\mathbf{r}}(N)$. Indeed, $p^{\mathbf{r}}(N) = \sum_{C \in 2^N \setminus \emptyset} \alpha^{\mathbf{r}}(C|N) p_C(N) = \sum_{C \in 2^N \setminus \emptyset} \alpha^{\mathbf{r}}(C|N) p_C(C) = \sum_{C \in 2^N \setminus \emptyset} \alpha^{\mathbf{r}}(C|N) v(C) = v^{\mathbf{r}}(N)$, where the last equality is just the definition of $v^{\mathbf{r}}$ (see Eq. (1)). It is also easy to show that no coalition $S$ can block $p^{\mathbf{r}}$. The second direction follows trivially, as any subgame $v|_C$ is a reliability extension of $v$ where $r_i = 1$ iff $i \in C$ and 0 otherwise. □

---

[2] We focus on *simple* base games. It is easy to extend the results to general (non-boolean) games, by setting a coalition's value in the reliability extension to its *expected* revenue after retaining only the surviving agents.

[3] Note that $p_C$ only set the payoff of agents in $C$. We therefore pad it with zeros, i.e. $p_C(i) = 0$ for all $i \notin C$.

## 5.1 Simple games

Even if the base game is simple, the reliability extension is typically not a simple game. Indeed, unless all the components of the probability vector **r** are either 0 or 1, in the reliability extension some coalitions are going to have a fractional value. We first discuss veto agents. In a simple game $i$ is a veto agent if it is criticcal in every coalition, i.e. if $i \notin C$ then $v(C) = 0$. In a non-simple game $i$ is a veto agent if no coalition has a strictly positive value without the agent.

**Lemma 2 (The Reliability Extension Preserves Veto Agents).** *Let $v$ be a simple monotone game used as a base game. If $i \in N$ is a veto agent in the base game $v$, then for any probability vector $\mathbf{r} = (r_1, \ldots, r_n)$, it holds that $i$ is a veto agent in $v^\mathbf{r}$.*

*Proof.* Let $i$ be a veto agent in $v$. Assume by contradiction that it is not a veto agent in $v^\mathbf{r}$, so for some coalition $C$ that $i \notin C$ we have $v^\mathbf{r}(C) > 0$. However, by definition of the reliability extension $v^\mathbf{r}(C) = \sum_{C' \subseteq C} \alpha^\mathbf{r}(C'|C) \cdot v(C')$, so at least one component in the sum must be strictly positive. Denote this component as $\alpha^\mathbf{r}(C^*|C) \cdot v(C^*)$. Thus we have $v(C^*) > 0$. By monotonicity, since $C^* \subseteq C$ then $0 < v(C^*) \leq v(C)$ in contradiction to $i$ being a veto agent. □

A known folklore theorem relates the core and veto agents in simple games. It states that in a simple game the core is empty if and only if the game has no veto agents. Further, in a simple game with veto agents, any imputation that allocates 0 to the non-veto agents and allocates $v(N)$ in total to the veto agents (in any way) is in the core.

**Corollary 2.** *Let $v$ be a simple monotone game to be used as a base game. If $v$ has a non-empty core then for any probability vector $\mathbf{r}$ the reliability extension $v^\mathbf{r}$ of $v$ has a non-empty core.*

*Proof.* First note that a simple game with a non-empty core is always convex. This is simply since the game must have a veto agent, and therefore every two winning coalitions intersect. Thus $v(S \cup T) + v(S \cap T) \geq v(S) + v(T)$ (there is actually an equality). Therefore, by Theorem 3, the core of $v^\mathbf{r}$ is non-empty. □

By Theorem 3 and Lemma 2, we can also find imputations in the core, when the core of the base game is non-empty. One simply allocates all the value of the game to the veto agents of the base game, in an arbitrary way. Since all veto agents must be contained in every subgame $v|_C$ with value greater than zero, this results in a stable imputation $p^\mathbf{r}$.

The folklore theorem also shows the converse regarding veto agents and the core: if a simple game has no veto agents then the core is empty. In the reliability extension of a game the core may be non-empty even when the base game has no veto agents.

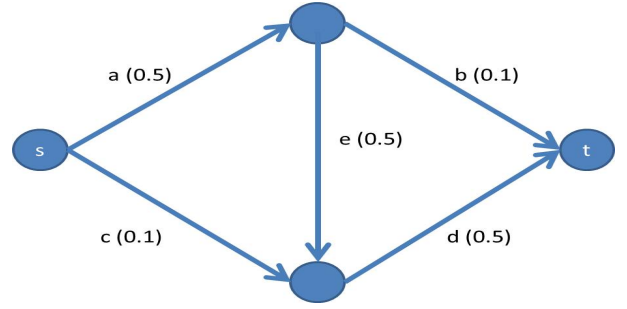

Figure 1: A network reliability game with an empty-core in the base game. There are 5 agents $N = \{a, b, c, d, e\}$. Survival probabilities are $r_a = r_d = r_e = 0.5$ and $r_b = r_c = 0.1$.

**Observation 1.** *There exist a simple game $v$ with no veto agents (and thus with an empty core) such that its reliability extension for some probability vector $(r_1, \ldots, r_n)$ has a non-empty core.*

*Proof.* Consider the network reliability game with five edges in Figure 1. It is easy to see that for a coalition to win in the base game it must contain at least one of the following edge subsets $\{a, b\}, \{c, d\}, \{a, e, d\}$. The base game has no veto agent - edges $a$, $b$ and $e$ are not veto agents as the coalition $\{c, d\}$ wins without them, and edges $c$, $d$ are not veto agents as the coalition $\{a, b\}$ wins without them. Thus the base connectivity game has an empty core.

It is easy to verify that the resulting reliability extension game for this probability vector has the following values: $v(\{a, b\}) = v(\{c, d\}) = 0.05$, $v(\{a, d, e\}) = 0.125$, $v(\{a, b, c, d\}) = 0.0975$, $v(\{a, b, c, d, e\}) = 0.19875$. For all the other edge subsets, their value is either 0, or the maximal value over the previous subsets that they contain. Consider the following imputation $r_a = r_c = 0$, $r_b = r_d = 0.05$ and $r_e = v(N) - r_b - r_d = 0.09875$. It is easy to verify that for every edge subset $C$ we have $p(C) = \sum_{i \in C} r_i \geq v(C)$ (to see that this holds, one needs only consider the previously discussed coalitions and verify the core constraint holds for them). Since this is a core imputation, the core is not empty and we have a simple game with an empty core whose reliability extension has a non-empty core.

Consider the same game where $e$ is eliminated from the base game. The base game still has an empty core. The extension of this game (with the same survival probabilities for $\{a, b, c, d\}$) also has an empty core. □

Observation 1 shows that although applying the reliability extension (for any probability vector) to a simple game can never eliminate the core if it exists, applying it may or may not make the core non-empty when the original game had an empty core. We have not been able to construct a polynomial algorithm for testing whether the core is empty for general reliability games.

### 5.2 Cores of Reliability Games With Few Agent Types and Survival Probabilities

We provide an algorithm for finding core imputations in reliability games with few survival probabilities and agent types. We first discuss typed-games. We say that two agents $i, j \in N$ are *equivalent* if for any coalition $C$ such that $i, j \notin C$ we have $v(C \cup \{i\}) = v(C \cup \{j\})$. Typed-games consider equivalence classes of agents. A typed game is a game with a set $T$ of agent types $(t_1, \ldots, t_k)$, where each agent $i \in N$ has a type $t(i) \in T$ and the value $v(C)$ of a coalition $C$ only depends on the types of the agents in $C$. Given a coalition $C$ we denote $q_j(C) = |\{i \in C | t(i) = t_j\}|$ (i.e. the number of agents of type $t_j$ in $C$), and $\#T(C) = (q_1(C), \ldots, q_k(C))$. In a typed-game we require that if for two coalitions $A, B$ we have $\#T(A) = \#T(B)$ then $v(A) = v(B)$. A coalition's value only depends on the quantities of agents of each type, so we can express the characteristic function as $v : (\mathbb{N} \cup \{0\})^k \to \mathbb{R}$, where $v(q_1(C), q_2(C), \ldots, q_k(C))$ stands for the value $v(C)$ of any coalition $C$ that contains $q_j$ agents of type $t_j$.

Our algorithm operates on a reliability extension game $v^\mathbf{r}$, where the base game $v$ is a typed-game where the number of different agents types is bounded by a *constant* $k$. We further assume that the survival probability vector $\mathbf{r}$ has a *constant* number $m$ of different probabilities. Thus, each $r_i \in R = \{r_1, r_2, \ldots, r_m\}$ (where $R$ is a set of $m$ different probability values). We call such games *limited reliability games*.

Given an agent $i$ in the reliability extension $v^\mathbf{r}$, we denote its type in the base game as $t_{base}(i)$. We further note that if two agents $i, j$ have the same type in the base game ($t_{base}(i) = t_{base}(j)$), and both have the same survival probability ($r_i = r_j$), then the two agents are equivalent in the reliability extension (as can be seen directly from the formula for $v^\mathbf{r}(C)$ in Definition 3). Both $k, m$ are constants, so w.l.o.g. we assume that agents of the same type also have the same survival probability, and are therefore equivalent (this only increases the number of types from $k$ to $mk$).

Consider a coalition $C$ such that $\#T(C) = (q_1, \ldots, q_k)$. We denote the probability that exactly $w$ of type $t_i$ would survive as $P_i(w, C)$. we have $P_i(w, C) = \binom{q_i(C)}{w}(r_i)^w(1-r_i)^{(q_i(C)-w)}$. We can then express the value of every coalition as follows:

$$v^\mathbf{r}(C) = \sum_{w_1=0}^{q_1(C)} \cdots \sum_{w_k=0}^{q_k(C)} \prod_{i=0}^{k} P_i(w_i, C) v(w_1, \ldots, w_k).$$

Since $k$ is constant and $q_i(C) \leq n$, we can compute $v(C)$ using this formula in time $O(n^k)$, which is polynomial in the number of agents.

We now turn to finding core imputations. The core can be defined using a simple feasibility linear program over the variables $p_1, \ldots, p_n$ that has a constraint for any coalition $C$, stating the $\sum_{i \in C} p_i \geq v(C)$. However, since the number of different coalitions is $2^n$, this program has an exponential number of constrains, so it is intractable to use it find core imputations. To avoid this problem, use the following well-known folklore theorem regarding cooperative games: if the core is non-empty then there exists a core imputation $\mathbf{p} = (p_1, \ldots, p_n)$ where for any two agents $i, j$ that are equivalent we have $p_i = p_j$. In other words, if the core is non-empty then it contains an imputation where all agents of the same type have the same payoff.

In a limited reliability game we have a constant number $k$ of agent types, so we can maintain a variable for each extended type. Denote the payment for an agent of *base* type $t_i$ by $p(i)$ (where $1 \leq i \leq k$). We now construct a feasibility linear program over the $p(i)$ variables. We say that a coalition is of profile $\mathbf{q} = (q_1, \ldots, q_k)$ if $\#T(C) = \mathbf{q}$ (i.e. if the coalition contains the specified amounts of agents of each extended type). There are $k$ types and $n$ agents, so the number of different coalition profiles is at most $n^k$, polynomial in the number of agents (as $k$ is *constant*). All payoffs of agents of the same type are identical, thus for every coalition $C$ of profile $\mathbf{q}$, $p(C) = \sum_{i=1}^{k} q_i p(i)$. Finally, we have the following core constraint for each coalition profile $\mathbf{q}$:

$$\sum_{i=1}^{k} q_i p(i) \geq v(q_1, \ldots, q_k).$$

We therefore have a linear program over $k$ variables, $p(1), \ldots, p(k)$ with a polynomial number of constrains. If we solve this program (e.g. using the ellipsoid method), we obtain a core imputation. If there is no solution, then the core is empty.

## 6 Conclusions and Related Work

We proposed the reliability extension of cooperative games, and discussed algorithms for solving this extension. Solving cooperative games has been well-studied and solutions have been offered for various representations [10, 6, 8] and even for games with uncertain payoffs [21]. To our knowledge, this is the first paper to consider computational issues in cooperative games with uncertain rewards due to independent failures. Although computing the Shapley value is hard in various domains, several approximation methods have been suggested [10, 9, 11]. Our sampling approach for approximating the Shapley value is similar to [3], but does not assume a method for computing the characteristic function of any coalition in polynomial time (as mentioned, for the network reliability game, computing a coalition's value is #P-hard). We also showed how to compute the core when the base game has veto agents and for limited reliability games, and noted important differences between the core of a simple game and the core of its extension. We used game theoretic solutions to find reasonable reward allocations. Other applications of such solutions include analyzing prediction markets [7] and measuring inconsistencies in probabilistic knowledge bases [20].

Some questions remain open. On the conceptual level, it is important to better characterize the conditions on the base game and survival probability that guaranty the non-emptiness of the core. On the algorithmic level, natural questions from computational game theory should be answered for this new rich family of games. What algorithms can compute the core of unrestricted reliability games? Can the Shapley value be exactly computed in restricted forms of these games? Can one tractably compute other solutions such as the kernel or least-core (see [15]) and the Cost of Stability [2, 17] in such games?

**Acknowledgment** The third author is partially supported by the Israel Science Foundation (grant number 1219/09) and by the Leon Recanati Fund of the Jerusalem school of business administration.